\documentclass[prl,english,preprintnumbers,amsmath,amssymb,nofootinbib,twocolumn,superscriptaddress]{revtex4}
\pdfoutput=1
\usepackage[latin1]{inputenc}
\usepackage{graphicx}
\usepackage{color}
\usepackage{bbm}
\usepackage{amssymb}
\usepackage{amsmath}
\usepackage{tabularx}
\usepackage{dsfont}
\usepackage{multirow}
\usepackage{dcolumn}

\def\0#1#2{\frac{#1}{#2}}

\def\s0#1#2{\mbox{\small{$ \frac{#1}{#2} $}}}

\newcommand{\beq}{\begin{equation}}
\newcommand{\eeq}{\end{equation}}
\newcommand{\bea}{\begin{eqnarray}}
\newcommand{\eea}{\end{eqnarray}}

\usepackage{babel}
\makeatother

\begin{document}

\title{Quantum anomaly and thermodynamics of one-dimensional \\ fermions with three-body interactions}

\author{J. E. Drut}
\affiliation{Department of Physics and Astronomy, University of North Carolina. Chapel Hill, North Carolina 27599-3255, USA}
\author{J. R. McKenney}
\affiliation{Department of Physics and Astronomy, University of North Carolina. Chapel Hill, North Carolina 27599-3255, USA}

\author{W. S. Daza}
\affiliation{Physics Department, University of Houston. Houston, Texas 77024-5005, USA}
\author{C. L. Lin}
\affiliation{Physics Department, University of Houston. Houston, Texas 77024-5005, USA}
\author{C. R. Ord\'o\~nez}
\affiliation{Physics Department, University of Houston. Houston, Texas 77024-5005, USA}

\begin{abstract}
We show that a system of three species of one-dimensional fermions, with an attractive three-body contact interaction, features a scale anomaly
directly related to the anomaly of two-dimensional fermions with two-body forces. We show, furthermore, that those two cases (and their multi-species generalizations)
are the only non-relativistic systems with contact interactions that display a scale anomaly. 
While the two-dimensional case is well-known and has been under study 
both experimentally and theoretically for years, the one-dimensional case presented here has remained unexplored.
For the latter, we calculate the impact of the anomaly on the equation of state, which appears through the generalization of Tan's contact for three-body forces, 
and determine the pressure at finite temperature. In addition, we show that the third-order virial coefficient is proportional to the second-order 
coefficient of the two-dimensional two-body case.
\end{abstract}

\date{\today}

\maketitle 

{\it Introduction.-}
The study of manifestations of scaling SO(2,1) anomalies in nonrelativistic systems has received 
considerable attention in recent years. Such anomalies appear when a symmetry is present at the classical level, 
but is broken by quantum fluctuations; the prime example in nonrelativistic physics is the two-dimensional (2D) Fermi gas with 
attractive contact interactions~\cite{ jackiw1991mab, PitaevskiRosch, Hofmann}. On the experimental side, ultracold-atom experiments have 
shed light on the thermodynamic, collective-mode, and transport properties of that 
2D system~\cite{Experiments2D2010Observation, 
Experiments2D2011Observation, 
Experiments2D2011RfSpectroscopy, 
Experiments2D2011DensityDistributionTrapped, 
Experiments2D2011Crossover2D3D, 
Experiments2D2012RfSpectraMolecules, 
Experiments2D2012Crossover2D3D,
Experiments2D2012Polarons, 
Experiments2D2012Viscosity,
ContactExperiment2D2012, 
Vale2Dcriteria,
Experiments2D2014, 
Experiments2D2015SpinImbalancedGas, 
Experiments2D2015PairCondensation,
Experiments2D2015BKTObservation} (see also~\cite{RanderiaPairingFlatLand,PieriDanceInDisk}).
On the theory side,
there have been multiple non-perturbative studies of basic ground-state ~\cite{Bertaina,ShiChiesaZhang,Gezerlis} and 
thermodynamic~\cite{LiuHuDrummond,Enss2D,ParishEtAl,AndersonDrut,BarthHofmann} 
quantities, and transport~\cite{ChafinSchaefer,EnssUrban,EnssShear}.
In particular, Ref.~\cite{UHetUNC2D} spelled out the relationship between these anomalies 
and the Tan contact for 2D fermion systems with two-body contact interactions, and put forward a computational 
framework to access the shift of the virial coefficients $\Delta b_n$, $n\ge 2$ using a Hubbard-Stratonovich transformation.

In this work, we show that a system of one-dimensional (1D) fermions with 
an attractive {\it three-body} contact interaction 
presents a scaling anomaly
of the same kind as that of the 2D case with two-body forces.
Naturally, the system is non-trivial only if at least three fermion species are present in the problem,
which implies straightforward results (e.g., the virial coefficients are non-trivial starting at third order)
as well as more challenging aspects (namely dealing with a three-body problem for any useful calculation).
While here we consider unpolarized distinguishable species (no mass asymmetry or population imbalance), 
generalizations to more species and asymmetric cases could and should also be studied.

{\it Hamiltonian.-}
The system is defined by the following Hamiltonian
\beq
\hat H = \!\!\!\!\sum_{s=1,2,3}\int dp \; \epsilon(p) \; \hat a_{s,p}^\dagger \hat a_{s,p} + g \int dx \; \hat n_1(x)  \hat n_2(x)  \hat n_3(x),
\eeq
where $\epsilon(p) = \frac{\hbar^2p^2}{2m}$.
Here, $\hat a_{s,p}^\dagger$ and $\hat a_{s,p}$ are the creation and annihilation operators
for particles of species $s$ and momentum $p$, and $\hat n_s(x)$ is the corresponding density
at position $x$. In what follows we will take $\hbar = k_B = m = 1$ (we will, however, show $m$
in the following equations in order to distinguish it from the total and reduced masses).
A crucial feature of this system is that, since the 1D density has units of inverse length, the bare coupling $g$ is dimensionless.
As we show below, however, the coupling runs non-trivially with the cutoff, and the physical coupling 
(dimensionally transmuted scale~\cite{Camblong}) 
is the binding energy of the three-body system.

{\it Renormalization and the three-body problem.-} 
As anticipated, in order to renormalize the problem we determine the connection between 
the bare coupling $g$ and the binding energy $\epsilon_B$ of the three-body system.
We will show here that the system forms such a three-body bound state at arbitrarily small $g$,
and we will do so by mapping our problem onto a 2D one-body problem interacting with an external
Dirac delta potential. That problem is of course what results from considering a two-body
problem with a two-body delta function interaction, when going to the center-of-mass frame.

The 1D three-particle Schr\"odinger equation for our system takes the form
\beq
\label{Eq:Sch3B}
\left[-\frac{\nabla^2_X}{2m} + 
g \delta(x_2 - x_1)\delta(x_3 - x_2)\right]
\psi(X)
=
E \psi(X)
\eeq
where we used the shorthand notation $X = (x_1,x_2,x_3)$ and
%
$\nabla^2_X = \frac{\partial^2}{\partial x_1^2}\!+\! \frac{\partial^2}{\partial x_2^2} \!+\! \frac{\partial^2}{\partial x_3^2}$.
%
One way to see the equivalence advertised above is already evident at this point: 
Eq.~(\ref{Eq:Sch3B}) corresponds to a 3D one-body problem (if we identify the coordinates by $(x,y,z) = (x_1,x_2,x_3)$) 
with an external line-like delta potential saturating at $x=y=z$. By symmetry, we may factorize such a 3D problem into a trivial part for the 
unrestricted motion parallel to the line, and a non-trivial part for the 2D motion perpendicular to the line (which sees a point-like delta potential).
As we show below, that factorization corresponds in the 1D problem to the center-of-mass (CM) and relative motions.

To be explicit, we proceed by separating the CM motion by the change of variables
$Q = \frac{1}{3} (x_1+x_2+x_3)$;  $q_1 = x_2  - x_1$; $q_2 = \frac{1}{{\sqrt{3}}} (x_1 + x_2 - 2 x_3)$.
Writing $\psi(X) = \Phi(Q)\phi(q_1,q_2)$, we obtain an equation for the CM motion, as usual,
\beq
\frac{-\nabla^2_Q}{2 {M}}\Phi(Q) = E_\text{CM} \Phi(Q),
\eeq
where $M = 3m$. For the relative coordinates $q_1$, $q_2$
[Note $q_2 = (2/\sqrt{3})(x_2 - x_3)$ when $q_1 = x_2 - x_1 = 0$],
\beq
\left[\frac{-\nabla^2_q}{2 \bar{m}}  + \tilde{g} \delta(q_1)\delta(q_2)\right]\phi(q_1, q_2) = E_r \phi(q_1, q_2),
\eeq
where $\bar m = m/2$ is the reduced mass, $\tilde{g} = (2/\sqrt{3})g$, is the effective coupling, $E_r$ is the energy of relative motion, and
%
$\nabla^2_q = \frac{\partial^2}{\partial q_1^2} + \frac{\partial^2}{\partial q_2^2}$,
%
which thus reduces the problem to that of a single particle in 2D with a delta function potential at the origin.
The problem is easily solved in momentum space, where one finds that the wavefunction takes the form
$\tilde{\phi}({\bf p}) \propto 1/({\bf p}^2 - E_r)$, and the binding energy $\epsilon_B = - E_r$ of the three-body bound state
(trimer) depends on the coupling as
\beq
\label{Eq:ContiuumEb}
\frac{\epsilon_B}{\Lambda^2} = e^{4\pi / \tilde{g}},
\eeq
where $\tilde g < 0$, and $\Lambda$ is a momentum cutoff that is required to regularize ultraviolet divergences.
Using the above relation, one identifies the trimer binding energy $\epsilon_B$ 
as the physical coupling, and as the emerging scale that breaks scale invariance.

It is noteworthy that for $n$-body contact interactions in $d$ dimensions, the 
units of the bare coupling are $L^{-2 + d(n-1)}$, such that there are {\it only} two physically relevant
solutions for which the coupling is dimensionless: $n=d=2$, i.e. the 2D case with a two-body 
interaction, and $n=3$, $d=1$, which is the 1D case shown here.

Below, we will use a lattice regularization of the problem to arrive at many-body results.
In that case, the relation between the bare lattice coupling $g_\text{lat}$ and the binding energy
is given implicitly by
\beq
\label{Eq:glatt}
\frac{1}{g_\text{lat}} = -\frac{1}{L^2} \sum_{\bf k} \frac{1}{\epsilon_{\bf k} + \epsilon_B},
\eeq
where $L = N_x \ell$ is the lattice size, $\ell$ is the lattice spacing, $\epsilon_{\bf k} = (k_1^2 + k_2^2 +k_3^2)/2$, 
${\bf k} = (2\pi/L)(n_1,n_2,n_3)$, and the sum covers $0 \leq |n_1 + n_2| \leq \Lambda$, with the constraint $n_1 + n_2 +n_3 = 0$ 
(i.e. vanishing total momentum).

{\it Results.-}
{\it Anomaly in the equation of state.} In truly scale invariant systems, such as noninteracting ones, the pressure $P$ may be written in terms
of the inverse temperature $\beta$ and the chemical potential $\mu$ as $P = \beta^{\alpha} f(\beta \mu)$,
where $\alpha = -d/2 - 1$ and $d$ is the number of spatial dimensions. The advantage of isolating the 
dependence on the dimensionful parameter $\beta$ is that one readily derives, using thermodynamic
identities and partial differentiation with respect to $\beta$ and $\mu$, the well-known result 
\beq
\label{Eq:ScaleInvEOS}
P = \frac{2}{d}\frac{E}{V},
\eeq
where $E$ is the total energy and $V$ is the $d$-dimensional volume.
In scale-anomalous systems like the one put forward here, the pressure acquires a second physical, dimensionless 
parameter via the anomaly, which we will write as $\beta \epsilon_B$, where $\epsilon_B$ is the binding energy
of the three-body problem described above. Thus, $P = \beta^{\alpha} f(\beta \mu, \beta \epsilon_B)$.
Following the same derivation outlined above, one can easily see that
\beq
\label{Eq:ScaleAnomEOS}
P - \frac{2}{d}\frac{E}{V} =  \frac{2}{d}\beta^{\alpha} \frac{\partial f}{\partial (\beta \epsilon_B)} \beta \epsilon_B = 
\frac{2}{d}\beta^{\alpha} \frac{\partial f}{\partial \ln (\beta \epsilon_B)},
\eeq
which shows that the emergence of the second parameter results in a contribution to the equation of state
that breaks the scale invariant result of Eq.~(\ref{Eq:ScaleInvEOS}).

{\it Anomaly as Tan's contact.}
Specializing to our case, the anomalous term in Eq.~(\ref{Eq:ScaleAnomEOS}) is proportional to a generalization of Tan's contact to
the case of 3-body forces. Indeed, since $\beta P V = \ln \mathcal Z$, where $V=L$ is the volume and 
$\mathcal Z = \text{Tr} \exp\left[- \beta (\hat H - \mu \hat N)\right]$ is the grand-canonical partition function, 
the only way in which $f$ can depend on $\epsilon_B$ is through the dimensionless bare coupling $g$ that 
appears in $\hat H$:
\beq
\frac{\partial f}{\partial \ln(\beta \epsilon_B)} =
\frac{\sqrt{\beta}}{L} \frac{\partial \ln \mathcal Z}{\partial g} \frac{\partial g}{\partial \ln(\beta \epsilon_B)},
\eeq
where

\beq
\frac{1}{\beta L}\frac{\partial \ln \mathcal Z}{\partial g} = - \langle \hat n_1  \hat n_2  \hat n_3 \rangle,
\eeq
and the angle brackets denote a thermal expectation value in the grand-canonical ensemble.
Thus, for our scale-anomalous 1D system
\beq
\label{Eq:ScaleAnomEOS2}
P - 2\frac{E}{L} =  {\mathcal C_3},
\eeq
where we have identified
\beq
\mathcal C_3 = 2 \frac{\partial P}{\partial \ln(\beta \epsilon_B)} = - 2\frac{\partial g}{\partial \ln(\beta \epsilon_B)} \langle \hat n_1  \hat n_2  \hat n_3 \rangle ,
\eeq
as the generalization of Tan's contact density for the case of three-body forces~\cite{TanContact,Valiente,ContactReview,WernerCastin}. Note that the dimensions of
the contact density are those of pressure or energy density, which in 1D amounts to $1/L^3$. Thus,
the contact factorizes into a three-body piece (the change in the bare coupling with the 
physical coupling) and a many-body piece (the expectation value of the triple density operator).
Note that, in the continuum, from Eq.~(\ref{Eq:ContiuumEb}) we find 
\beq
\frac{\partial g}{\partial \ln (\beta \epsilon_B)} = -\frac{1}{2\pi \sqrt{3}} g^2.
\eeq
On the lattice, using the relationship between $g_\text{lat}$ and $\epsilon_B$,
\beq
\frac{\partial g_\text{lat}}{\partial \ln (\beta \epsilon_B)} = - g^2_\text{lat} \frac{1}{L^2} \sum_{\bf k} \frac{\epsilon_B}{(\epsilon_{\bf k} + \epsilon_B)^2}.
\eeq
where the sum is constrained in the same way as that of Eq.~(\ref{Eq:glatt}).


{\it Virial coefficients and high-temperature thermodynamics.}
Because the system proposed here contains no two-body forces, the coefficients $b_n$ of the virial expansion
are identical to those of the non-interacting case up to second order: $b_1 = 1$; $b_2 = b_2^{(0)}$,
where in general $b_n^{(0)} = (-1)^{n+1}/n^{3/2}$.
The third-order coefficient $b_3$ and above, however, are directly affected by the anomaly. Indeed, if the interacting $n$-body partition function is $Q_n$,
the first nontrivial one in our system is $Q_3$. Therefore,
\beq
\Delta b_3 \equiv b_3 - b_3^{(0)} = \frac{\Delta Q_3}{Q_1},
\eeq
where $b_3^{(0)}$ is the noninteracting third order virial coefficient, and we have used the definition
$b_3 \equiv {Q_3}/{Q_1} - Q_2 + Q_1^2/3$ together with the fact that $Q_1$ and $Q_2$ are unaffected
by the three-body interaction. Moreover, $\Delta Q_3 = \Delta Q_{1,1,1}$, where
$Q_{n_1,n_2,n_3}$ is the partition function of the system with $n_j$ particles of species $j$.
In translation-invariant systems it is always possible to factor out the center-of-mass motion,
such that $\Delta Q_3 = Q_{3}^\text{CM} \Delta Q_{1,1,1}^\text{rel}$, where $Q_{3}^\text{CM} = \sqrt{3} L/\lambda_T$
and $\lambda_T = \sqrt{2\pi\beta}$.
Similarly, the two-body 2D case satisfies
\beq
\Delta b_2^\text{2D} \equiv b^\text{2D}_2 - b_2^{(0),\text{2D}} = 
\frac{\Delta Q^\text{2D}_2}{Q^\text{2D}_1} = \frac{Q_{2}^\text{CM,2D} \Delta Q^\text{rel, 2D}_{1,1}}{Q^\text{2D}_1},
\eeq
where $Q_{2}^\text{CM,2D} = 2 L^2/\lambda_T^2$.
Since we showed above that the relative motion of the three-body 1D problem is captured by the dynamics 
of the two-body problem in 2D, we have (at fixed $\beta \epsilon_B$), $\Delta Q_{1,1,1}^\text{rel} = \Delta Q_{1,1}^\text{rel,2D}$. 
Thus, putting together the above equations we arrive at
\beq
\Delta b_3 = \frac{Q_{3}^\text{CM}}{Q_1}\frac{Q^\text{2D}_1}{Q_{2}^\text{CM,2D}}\Delta b_2^\text{2D} = \frac{\Delta b_2^\text{2D}}{\sqrt{3}},
\eeq
where we used the expressions for the 1D, three-flavor single-particle partition function $Q_1 = 3 L/\lambda_T$,
and the 2D, two-flavor analogue $Q^\text{2D}_1 = 2 L^2/\lambda^2_T$. It is important to note that the factor of $1/\sqrt{3}$
in the above equation relating $\Delta b_3$ and $\Delta b_2^\text{2D}$ is unrelated to the factor of $2/\sqrt{3}$ that
appears in the relationship between $g$ and $\tilde g$.

From the above considerations we obtain the high-temperature (strictly speaking low-fugacity) behavior
of the pressure and Tan's contact using the corresponding virial expansions, namely
\bea
\beta L (P - P_0) &=& Q_1 \sum_{k=1}^{\infty} \Delta b_k z^k \\
\beta L {\mathcal C_3} &=& Q_1 \sum_{k=1}^{\infty} c_k z^k,
\eea
where $c_k = 2\partial b_k / \partial \ln (\beta \epsilon_B)$.

In addition, the relationship between $\Delta b_2^\text{2D}$ and $\Delta b_3$ yields the thermodynamics
of the three-body problem, since the change in the corresponding partition function is
\beq
Q_3 - Q_3^{(0)} = Q_1 \Delta b_3,
\eeq
where $Q_3^{(0)} = (Q_1/3)^3$.

{\it Toward the many-body properties.}
Despite the close connection between the 1D and 2D problems explained above,
in particular at the few-body level, the many-body properties certainly differ (e.g. we 
expect a superfluid transition in the 2D case, while no such behavior would appear in 
our 1D system). To provide a first look at the thermodynamics, we present here 
perturbative results for the pressure at finite temperature and density.

\begin{figure}[h]
\begin{center}
	\includegraphics[scale=0.7]{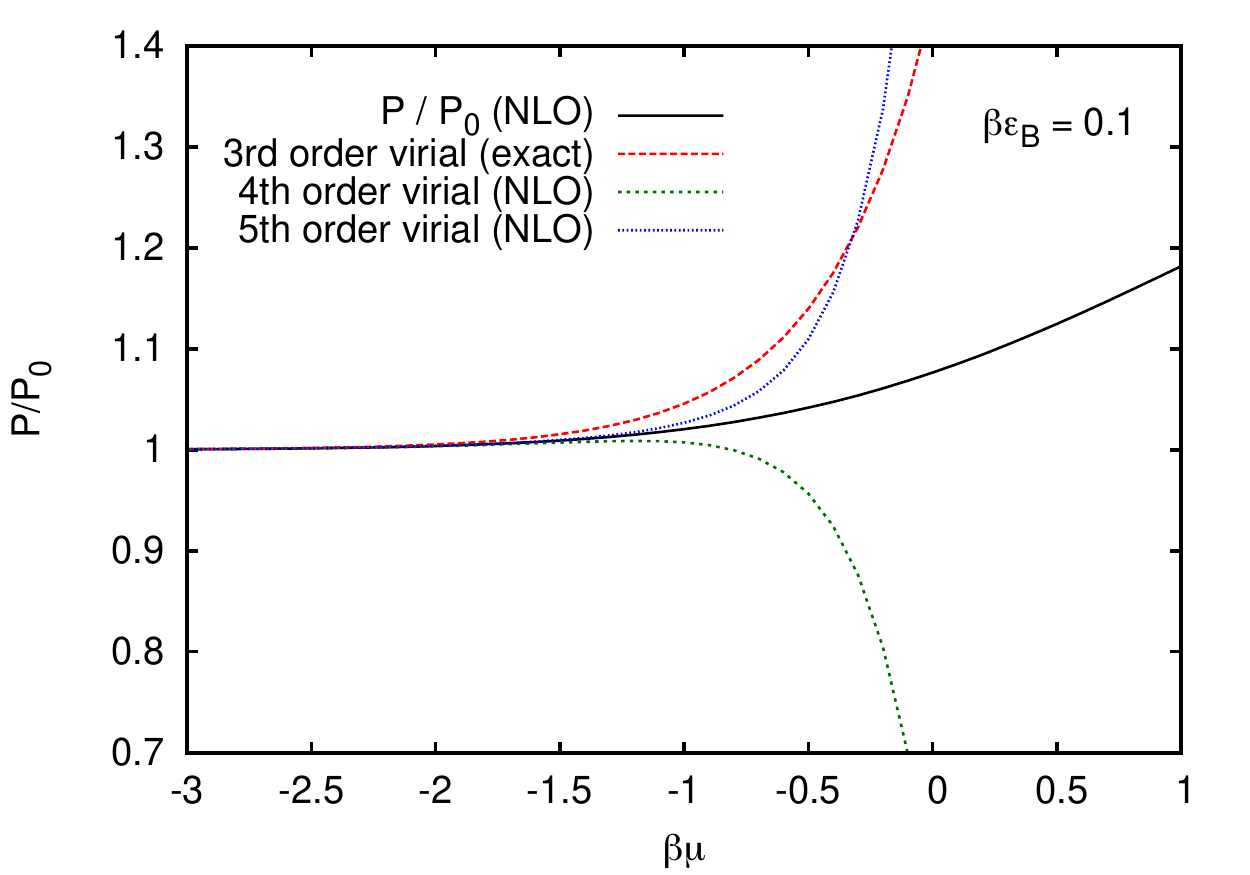}
	\caption{\label{Fig:Pressure} Solid line: Pressure, in units of of the noninteracting counterpart, of the 1D anomalous 
	many-body system at $\beta \epsilon_B = 0.1$, as a function of $\beta\mu$. The various virial-expansion results show the
	exact contribution up to third order ($\Delta b_3$ was used to tune the lattice coupling), and perturbative results up to fourth and fifth orders from the
	evaluation of the pressure at next-to-leading order in lattice perturbation theory.}
\end{center}
\end{figure}

To obtain those results, we put the system on a spacetime lattice and use a Hubbard-Stratonovich
transformation to represent the interaction. Specifically, we write for a given point in spacetime,
\beq
e^{-\tau g_\text{lat} \hat n_1  \hat n_2  \hat n_3} = \int_\Gamma \frac{d\sigma}{3\pi} \prod_{j=1,2,3}(1 + B \hat n_j f(\sigma)),
\eeq
where $\Gamma = [{-3\pi/2},{3\pi/2}]$, $f(\sigma) = e^{i 2 \sigma / 3} \cos^{2}\sigma$, $B^3 = C(e^{-\tau g_\text{lat}} - 1)$
with $C = 64/15$,  $g_\text{lat}$ is the lattice bare coupling, and $\tau$ is the temporal lattice spacing.
Using this transformation, we may write the partition function as
\beq
\mathcal Z = \int \mathcal D \sigma \; {\det}^3 M[\sigma],
\eeq
where the matrix $M[\sigma]$ is the usual fermion matrix encoding the dynamics and input parameters, including the fugacity $z$.
One may use this formulation of the many-body problem to carry out Monte Carlo calculations~\cite{Drut:2012md,QMCReviews}; 
however, the imaginary parts of $f(\sigma)$
imply that there would be a so-called phase problem. Instead, for such calculations one should resort to fixed-node methods, which
we will carry out elsewhere. Here, we evaluate the pressure at next-to-leading order in perturbation theory, as we outline next.
Expanding the effective action $S[\sigma] = -3 \ln \det M[\sigma]$ in powers of $B$, along the lines of the work of Ref.~\cite{PRDAndrew}, 
and keeping only the leading contribution, we obtain
\beq
\ln (\mathcal Z/\mathcal Z_0) = N_\tau N_x \ln \left[ \int_\Gamma \frac{d \sigma}{3\pi} \exp( 3 K B f(\sigma) ) \right],
\eeq
where
\beq
K = \frac{1}{N_x}\sum_{p}\frac{z e^{-\beta p^2/2}}{1 + z e^{-\beta p^2/2}}, 
\eeq
all of which is evaluated numerically.
To renormalize the coupling, we compute the change in the third-order virial coefficient in our approximation 
and tune $g_\text{lat}$ to match the exact $\Delta b_3$ derived above, and thus obtain the physical
coupling $\beta \epsilon_B$.
Using $\ln (\mathcal Z/\mathcal Z_0) = \beta V(P- P_0)$, in Fig.~\ref{Fig:Pressure} we show $P/P_0$ as a function of $\beta \mu$
for $\beta \epsilon_B = 0.1$, alongside the virial expansion, to illustrate the shape of the equation of state of this system.
The increase $P/P_0$ as a function of $\beta \mu$ is characteristic of systems with attractive interactions (see e.g.~\cite{EoS1D, AndersonDrut}).

Due to the formation of three-body bound states at vanishingly small attractive coupling, 
we expect to have an effective description in terms of composite fermions, i.e. trimers at strong coupling. 
As the trimer states become localized with increased coupling, 
Pauli exclusion dictates that their interaction should be repulsive. Thus, we expect a behavior that is
rather different from the fermion-boson crossover phenomenon featured in 2D; instead,
we expect a fermion--trimer crossover, where both ends are of fermionic character.
For weak attraction, where the trimers are loosely bound, the
trimer-trimer interaction may be attractive, which would result in trimer pairing.
At fixed $\ln (\epsilon_B/\epsilon_F)$, there should exist a crossover temperature $T^*$ above which
the trimers break into unbound fermions.
\begin{figure}[t]
\begin{center}
	\includegraphics[scale=0.315]{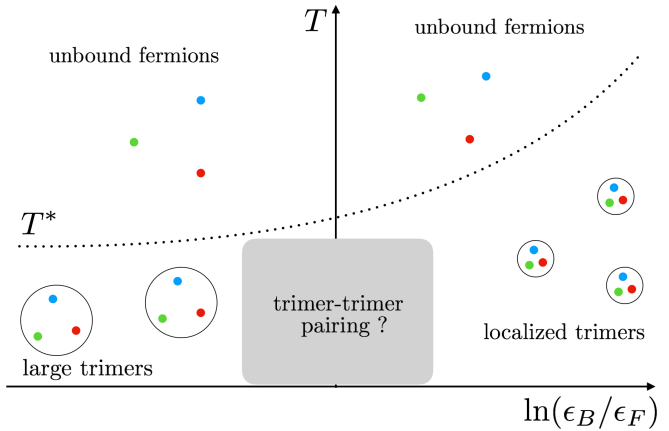}
	\caption{\label{Fig:Diagram} Conjectured many-body behavior of the 1D anomalous system as a function of temperature $T$
	and dimensionless coupling $\ln(\epsilon_B / \epsilon_F)$, where $\epsilon_F$ is the Fermi energy. At low temperatures,
	the system forms large trimers (at weak coupling) and localized repulsive trimers (point-like
	identical fermions, at strong coupling). Whether the attractive interaction is enough to overcome Pauli exclusion and 
	lead to trimer-trimer pairing, it remains an open question. At high enough temperatures (crossover pictured here as $T^*$), 
	the effective description should be in terms of unbound fermions with residual three-body correlations.}
\end{center}
\end{figure}
Finally, the more general situation where the particle population is asymmetric, e.g.
majority or minority type 1 and equal population of types 2 and 3, or all different, may lead
to a variety of situations (e.g. fermion-mediated attractive interaction between dimers),
to be explored elsewhere.

{\it Summary and Conclusions.-}
We have shown that a system of 1D fermions with 
an attractive three-body contact interaction features a scale invariance which, while
present at the classical level (the coupling $g$ is dimensionless), is broken by
quantum fluctuations which generate a three-body bound state at arbitrarily small couplings. 
To show it, we mapped the three-body 1D problem to a two-body 2D problem 
(or, rather, both are mapped onto the same one-body 2D problem
in a Dirac delta potential). Thus, this system presents 
a scale anomaly in a remarkable way: it lives in 1D but it is locally 
(around any region where particles scatter) like its 2D two-body counterpart,
which is reminiscent of the concept of holography. 
We have shown that the anomaly is directly related to Tan's contact,
which introduces a change in the equation of state in a way that is essentially
identical to that of the 2D case. In addition, we have shown that the third order virial
coefficient of our 1D system is proportional to the second-order coefficient of the 2D system.
Finally, we provide an initial look into the many-body properties with a perturbative calculation 
of the pressure, and conjecture an overall picture of the physics of the system in the 
temperature-coupling plane.

We acknowledge useful discussions with L. Dolan, A.~C. Loheac, and C. R. Shill.
This work was supported by the U.S. National Science
Foundation under Grant No. PHY1452635 (Computational Physics Program).
This work was supported in part by the US Army Research Office Grant No. W911NF-15-1-0445.


\end{document}